# Ultrahigh interfacial thermal conductance for cooling gallium oxide electronics using cubic boron arsenide

Wenjiang Zhou[1,2], Nianjie Liang[1], Wei Xiao[1], Zhaofei Tong[3], Fei Tian[3], and Bai Song[1,4,5*]

[1]*Department of Energy and Resources Engineering, Peking University, Beijing 100871, China.*

[2]*School of Advanced Engineering, Great Bay University, Dongguan 523000, China.*

[3]*School of Materials Science and Engineering, Sun Yat-sen University, Guangzhou, Guangdong 510006, China.*

[4]*Department of Advanced Manufacturing and Robotics, Peking University, Beijing 100871, China.*

[5]*National Key Laboratory of Advanced MicroNanoManufacture Technology, Beijing 100871, China*

[*]Author to whom correspondence should be addressed; e-mail: songbai@pku.edu.cn (B. Song)

**ABSTRACT**:

Gallium oxide ($Ga_2O_3$) has attracted significant interest for its unique potential especially in power electronics. However, its low and anisotropic thermal conductivity poses a major challenge for heat dissipation. Here, we explore an effective cooling strategy centering on the heterogeneous integration of $β$-$Ga_2O_3$ devices with cubic boron arsenide (cBAs), an emerging material with an ultrahigh thermal conductivity $κ$ of ~1300 $Wm^{-1}K^{-1}$. Machine-learned potentials for representative $β$-$Ga_2O_3$/cBAs interfaces are trained, enabling accurate and efficient calculation of the interfacial thermal conductance $G$ via nonequilibrium molecular dynamics. At 300 K, remarkable $G$ values of 749±33 $MWm^{-2}K^{-1}$ and 824±35 $MWm^{-2}K^{-1}$ are predicted for Ga-As and O-B bonding across the interface, respectively, which are primarily attributed to the well-matched phonon density of states considering the similar Debye temperatures of $β$-$Ga_2O_3$ and cBAs. Moreover, finite-element simulations directly show a notable device temperature reduction when comparing cBAs with other substrates. The simultaneously ultrahigh $κ$ and $G$ highlight cBAs as an ideal substrate for $Ga_2O_3$ electronics.

**KEYWORDS**: Gallium oxide; Cubic boron arsenide; Interfacial thermal conductance; Machine-learned molecular dynamics

Gallium oxide (Ga$_2$O$_3$) has garnered considerable attention for its potential use in power and radio-frequency electronics [1, 2], solar-blind UV detectors [3], and gas sensing devices [4]. Among the six crystalline phases of Ga$_2$O$_3$ [5, 6], $\beta$-Ga$_2$O$_3$ is particularly notable for combining a suite of exceptional properties, including thermal and chemical stability, ultrawide bandgap (~4.9 eV), simultaneously high intrinsic electron mobility (~300 cm$^2$V$^{-1}$s$^{-1}$) and critical electrical field (~8 MV/cm) which lead to a remarkable Baliga figure of merit up to 3400 times that of silicon [7-9], and also ultrahigh radiation resistance [10]. In addition, the low cost for manufacturing $\beta$-Ga$_2$O$_3$ further contributes to its rising prominence in the semiconductor industry. However, despite all these advantages, the rather low and anisotropic thermal conductivity of $\beta$-Ga$_2$O$_3$ (around 10 to 30 Wm$^{-1}$K$^{-1}$) [11-16] presents a major obstacle to effective heat dissipation, which substantially limits device performance and reliability.

Thermal management of electronic devices inevitably involves heat conduction through multiple materials and interfaces, as illustrated in Figure 1a. In homoepitaxial Ga$_2$O$_3$ devices, previous modeling predicted that the junction temperature would rise by approximately 1500 °C under a target power density of 10 W/mm [17]. To address this formidable issue, heterogeneous integration of Ga$_2$O$_3$ devices with substrates made of high thermal conductivity ($\kappa$) materials has emerged as a promising strategy to enhance heat dissipation [18, 19]. Among various high-$\kappa$ materials, cubic boron arsenide (cBAs) was recently revealed to feature an ultrahigh $\kappa$ of about 1300 Wm$^{-1}$K$^{-1}$ [20-22]. Remarkably, despite this outstanding thermal conductivity, cBAs has a rather low Debye temperature of about 700 K [23] which is strikingly close to that of $\beta$-Ga$_2$O$_3$ (~738 K) [13]. The similar Debye temperatures of $\beta$-Ga$_2$O$_3$ and cBAs may lead to a favorable match of their phonon density of states, which in turn suggests the possibility of achieving a high interfacial thermal conductance. If confirmed, this would render cBAs a potentially ideal substrate for cooling $\beta$-Ga$_2$O$_3$ devices, especially compared to other high-$\kappa$ materials such as silicon carbide (3C-SiC), cubic boron nitride (cBN), and diamond, all of which have notably higher Debye temperatures (Figure 1b). However, to the best of our knowledge, no experimental measurements or theoretical calculations of thermal transport across any $\beta$-Ga$_2$O$_3$/cBAs interface have been reported so far. While experiments may have been largely limited by challenges in forming high-quality $\beta$-Ga$_2$O$_3$/cBAs heterostructures, simulations are often impeded by the limited accuracy of empirical interatomic potentials [24].

In this Letter, we develop machine-learned potentials (MLPs) for representative $\beta$-Ga$_2$O$_3$/cBAs heterostructures and then compute the temperature-dependent interfacial thermal conductance ($G$) based on nonequilibrium molecular dynamics (NEMD) simulations. The NEMD calculations yield exceptionally high $G$ values which reach 749±33 MWm$^{-2}$K$^{-1}$ and 824±35 MWm$^{-2}$K$^{-1}$ at room temperature for Ga-As and O-B bonding, respectively, across the $\beta$-Ga$_2$O$_3$(010)/cBAs(111) interface. These ultrahigh interfacial thermal conductances are primarily attributed to the similar Debye temperatures of $\beta$-Ga$_2$O$_3$ and cBAs and their well-matched phonon density of states. In addition to the atomistic simulations, we further perform device-level thermal modeling by employing the finite-element method, which allows for a direct comparison of cBAs with other high-$\kappa$ substrates in terms of their cooling performance.

To begin with, we examine the phonon density of states (DOS) of $\beta$-Ga$_2$O$_3$ and two prototypical ultrahigh-$\kappa$ materials including cBAs and diamond, which are calculated using the density functional theory (DFT) as implemented in the Vienna Ab initio Simulation Package (VASP) [25, 26]. Our calculations use the primitive cells for cBAs and diamond, and a conventional cell for $\beta$-Ga$_2$O$_3$. The finite-displacement (FD) method as implemented in the Phonopy package [27] is employed for lattice-dynamics calculations. Detailed computational parameters are presented in Table S1 [28].

As shown in the inset of Figure 1b, compared to diamond, the normalized phonon DOS of cBAs exhibits a much better alignment with that of $\beta$-Ga$_2$O$_3$, particularly below 10 THz where acoustic phonons dominate heat transfer. To quantitatively assess the matching of phonon DOS between two materials, an overlap factor $S$ is defined as [29, 30]:

$$S = \frac{\int_0^\infty D_1(\omega)D_2(\omega)d\omega}{\sqrt{\left(\int_0^\infty D_1(\omega)^2 d\omega\right)\left(\int_0^\infty D_2(\omega)^2 d\omega\right)}}, \tag{1}$$

where $\omega$ represents the phonon frequency, and $D_1$ and $D_2$ are the phonon DOS. For $\beta$-Ga$_2$O$_3$ and diamond, the calculated $S$ is 0.15. In comparison, this value increases markedly to 0.57 when $\beta$-Ga$_2$O$_3$ is paired with cBAs instead. The higher degree of phonon DOS match facilitates more efficient phonon transmission at the interface, which is generally expected to result in a higher $G$ value [31-34].

To reliably quantify the thermal conductances of various $\beta$-Ga$_2$O$_3$/cBAs interfaces via NEMD simulations, it is essential to reconstruct the potential energy surface (PES) with high accuracy. To this end, we prepare an *ab initio* dataset and train a MLP model based on the neuroevolution potential (NEP) method [35]. This framework is readily available in the Graphics Processing Units Molecular Dynamics (GPUMD) package [36, 37], and has been widely used in the study of thermal transport [35, 36, 38, 39].

Figure 2a-c demonstrates the training configurations including supercells of bulk $\beta$-Ga$_2$O$_3$ and cBAs crystals (only the conventional unit cells are shown for clarity), as well as their heterostructure containing a $\beta$-Ga$_2$O$_3$/cBAs interface. For $\beta$-Ga$_2$O$_3$, the space group is $C2/m$ and the relaxed lattice parameters are $a = 12.46$ Å, $b = 3.09$ Å, and $c = 5.88$ Å with $\beta = 103.68°$. This low-symmetry lattice results in highly anisotropic thermal transport. For example, at 300 K, the thermal conductivity values along the [100], [010], and [001] directions were measured to be 9.5 Wm$^{-1}$K$^{-1}$, 22.5 Wm$^{-1}$K$^{-1}$, and 13.3 Wm$^{-1}$K$^{-1}$, respectively [12]. For cBAs, the space group is $F\bar{4}3m$ and the relaxed lattice parameter is $a = 4.82$ Å. So, there is clearly a large lattice mismatch between $\beta$-Ga$_2$O$_3$ and cBAs.

To construct the heterostructure, the unit cells need to be expanded for better alignment across the interface. This notably increases both the complexity and computational cost of generating a suitable training dataset for an arbitrary $\beta$-Ga$_2$O$_3$/cBAs interface using DFT calculations. Considering the highest thermal conductivity of $\beta$-Ga$_2$O$_3$ along [010] direction and its close relevance to practical applications [40-42], we focus on integrating the (010) plane of $\beta$-Ga$_2$O$_3$ with the (100) and (111) planes of cBAs. There are four types of atomic bonding configurations across the interface: Ga-B, Ga-As, O-B, and O-As. In order to maintain electrical neutrality and form a stable interface, it is crucial for the bonding atoms to have opposite electric charges [43]. Therefore, this work primarily focuses on the case of Ga-As and O-B. To sample these two bonds in one single configuration, no vacuum layer is applied along the $z$-direction, as shown in Figure S1 [28]. More details regarding the preparation of the training dataset, DFT calculations, and the training processes can be found in our previous work [44] and the Supplemental Material [28].

In Figure 2d-f, we show the parity plots for the atomic energy, forces, and virial stresses of the training and test dataset, respectively. In the training dataset, the root-mean-square errors (RMSEs) of the energy, forces, and virial stresses are 2.17 meV/atom, 151.27 meV/Å, and 14.18 meV/atom. For the test dataset, the corresponding RMSEs are 1.80 meV/atom, 146.64 meV/Å, and 12.88 meV/atom, respectively. We also display the error histograms for the training dataset in the insets of Figure 2d-f. The error histograms follow a typical Gaussian distribution, as previously observed in Refs. [45, 46]. Similar results for the test dataset are plotted in Figure S2 of the Supplemental Material [28]. To further confirm the accuracy of our NEP model, we compute the phonon dispersions for $\beta$-Ga$_2$O$_3$ and cBAs using the FD methods, as illustrated in Figure S3 [28]. The good agreement between DFT calculations and the NEP predictions demonstrates the high accuracy of our model. In addition, the temperature in the $NVT$ ensemble for the $\beta$-Ga$_2$O$_3$/cBAs heterostructure can be effectively maintained at the target value for a long time, at least on the order nanosecond as shown in Figure S4 [28].

With the NEP model ready, we proceed to perform NEMD simulations using the GPUMD package [37]. The simulation models are schematically illustrated in Figure S5 [28]. More calculation details can be found in the Supplemental Material [28]. In Figure 3a, we show an example of the steady-state temperature distribution across the $\beta$-Ga$_2$O$_3$(010)/cBAs(111) interface with O-B bonding. With linear fittings applied to within the $\beta$-Ga$_2$O$_3$ and cBAs regions, the temperatures at the two sides of the interface can be extrapolated, which differ by $\Delta T$ = 10.8 K [47, 48]. The heat flux in the NEMD simulations is calculated using the formula $J = |dQ/dt|\, A^{-1}$, where $A$ is the cross-sectional area and $|dQ/dt|$ is the energy transfer rate. As plotted in the inset of Figure 3a, the accumulated energy extracted from the heat source and added to the sink are both 1.19 keV/ns at 300 K, demonstrating excellent energy conservation in the NEMD simulations. The temperature and energy profiles for Ga-As bonding and also the $\beta$-Ga$_2$O$_3$(010)/cBAs(100) interface (both Ga-As and O-B bonding) are provided in Figure S6 [28]. Finally, the interfacial thermal conductance is calculated as $G = J/\Delta T$.

The NEMD-calculated $G$ as a function of temperature is plotted in Figure 3b. Indeed, due to the similar Debye temperatures and well-matched phonon DOS, exceptionally high $G$ values are obtained for all the $\beta$-Ga$_2$O$_3$/cBAs interfaces. At room temperature, $G$ reaches 389±3 MWm$^{-2}$K$^{-1}$ and 531±10 MWm$^{-2}$K$^{-1}$ for the $\beta$-Ga$_2$O$_3$(010)/cBAs(100) interface with Ga-As and O-B bonding, respectively. These values rise further to 749±33 MWm$^{-2}$K$^{-1}$ and 824±35 MWm$^{-2}$K$^{-1}$ for the $\beta$-Ga$_2$O$_3$(010)/cBAs(111) interface. As temperature rises, $G$ further increases, primarily due to the contribution from inelastic phonon interactions [49]. To understand the higher $G$ for O-B bonding, we look into the electronegativity (EN) values of Ga, O, B, and As, which are 1.81, 3.44, 2.04, and 2.18, respectively [50]. Therefore, the EN difference is 0.37 between Ga and As, and 1.4 for O and B. The larger EN difference leads to stronger bonding and higher binding energy which can be calculated as $E_\text{b} = (E_\text{AB} - E_\text{A} - E_\text{B})/N$. Here, $E_\text{AB}$ is the total energy of the heterostructure, $E_\text{A}$ and $E_\text{B}$ represent the energy of A and B before bonding, and $N$ is the total number of atoms. Taking the $\beta$-Ga$_2$O$_3$(010)/cBAs(111) interface as an example, the absolute binding energy for Ga-As is 77 meV/atom, which is significantly lower than the 189 meV/atom for the interface with O-B bonding.

To gain deeper physical insights, we calculate the spectrally decomposed $G$ [51, 52]. To this end, the virial-velocity correlation function in the non-equilibrium state is expressed as:

$$\mathbf{K}(t) = \sum_i \langle \mathbf{w}_i(0) \cdot \mathbf{v}_i(t) \rangle. \tag{2}$$

Here, $\mathbf{w}_i$ and $\mathbf{v}_i$ are the per-atom virial and velocity of atom $i$, respectively. By performing the Fourier transform of $\mathbf{K}(t)$, the spectrally decomposed $G$ can be obtained as:

$$G(\omega) = \frac{2 \int_{-\infty}^{+\infty} dt e^{i\omega t} \mathbf{K}(t)}{V \Delta T}, \tag{3}$$

where $V$ is the volume. The calculated spectral $G$ at 300 K is plotted in Figure 3c, the integral of which closely matches the total $G$ in Figure 3b. Phonons with frequencies below 10 THz contribute approximately 92% of the total $G$, indicating that the acoustic phonons are the main heat carriers across the $\beta$-Ga$_2$O$_3$/cBAs interface [53]. The absence of contributions in the 10~17 THz spectral range is attributed to the acoustic-optical (*ao*) phonon bandgap in cBAs, which makes it hard to satisfy energy conservation. Moreover, optical phonons at around 19 THz contribute approximately 5% of the total $G$. These observations are consistent with the position-dependent phonon DOS in the vicinity of the interface, as plotted in Figure 3d.

In addition to the bulk phonons, there can also be various interfacial vibrational states which contribute non-trivially to thermal transport [54, 55]. To extract the interfacial modes, we minimize the difference between the calculated DOS at the interface and the linear combination of the bulk DOS [56], as illustrated in Figure 4a-b. The bulk DOS of cBAs and $\beta$-Ga$_2$O$_3$ are acquired from regions located 10 nm away from the interface. As shown in Figure 4c-d, both the cBAs and $\beta$-Ga$_2$O$_3$ sides exhibit large positive residuals in the spectral range of 4-8 THz. This indicates the presence of interfacial modes that bridge the two sides via inelastic scattering and provide additional pathways for heat transport across the interface [57].

Compared to Ga-As bonding, interfaces with O-B bonding exhibit a higher density of interfacial modes which is consistent with the higher $G$.

Further, we compare the room-temperature $G$ values for the interfaces between $\beta$-Ga$_2$O$_3$ and various high-$\kappa$ substrates, as shown in Figure 5a. For the combination of $\beta$-Ga$_2$O$_3$ and diamond, Chen *et al.* measured a $G$ value of about 179 MWm$^{-2}$K$^{-1}$ [58] for an interface formed via atomic layer deposition, while a much lower value of 17 MWm$^{-2}$K$^{-1}$ [59] was reported for a van der Waals bonded interface. NEMD simulations based on an empirical potential and a MLP model predict the $G$ values of around 20 MWm$^{-2}$K$^{-1}$ [24] and 400 MWm$^{-2}$K$^{-1}$ [60], respectively. For the $\beta$-Ga$_2$O$_3$/3C-SiC interfaces, experimental measurements following surface-activated bonding and annealing at 1000 °C yielded a $G$ value of about 244 MWm$^{-2}$K$^{-1}$ [61]. NEMD simulations for the $\beta$-Ga$_2$O$_3$/Si interfaces based on an empirical potential predict a $G$ value of about 250 MWm$^{-2}$K$^{-1}$ [62]. These comparisons highlight the ultrahigh $G$ of approximately 800 MWm$^{-2}$K$^{-1}$ achieved at the $\beta$-Ga$_2$O$_3$/cBAs interfaces.

Finally, we perform device-level thermal modeling of a simplified $\beta$-Ga$_2$O$_3$ power device by employing the finite-element method (FEM), which enables a direct comparison of cBAs with other high-$\kappa$ substrates in terms of their cooling performance. The simulated model is provided in Figure S5 of the Supplemental Material [28]. A high power density of about 10 W/mm and three $\beta$-Ga$_2$O$_3$ layer thicknesses (100 nm, 300 nm, and 500 nm) are considered [17]. Additional key parameters used for the FEM simulations are provided in Table S3 [28]. Compared to the extreme heating in the homoepitaxial $\beta$-Ga$_2$O$_3$ devices, the heterogeneous integration with high-$\kappa$ substrates significantly reduces the maximum temperature, lowering it by approximately 1200°C. For the case of 300 nm-thick $\beta$-Ga$_2$O$_3$, the cBAs substrate reduces the maximum device temperature by 4°C, 30°C, and 117°C compared to diamond, 3C-SiC, and Si, respectively. Across all the three device-layer thicknesses, cBAs consistently demonstrates better heat dissipation performance. Notably, in 100 nm and 300 nm $\beta$-Ga$_2$O$_3$/cBAs devices, a safe operating temperature below the threshold of 225°C can be maintained at the power density of 10 W/mm.

In summary, we explore the efficient cooling of $\beta$-$Ga_2O_3$ devices via heterogeneous integration with cBAs based on a suite of computational approaches including the density functional theory, machined-learned potentials, non-equilibrium molecular dynamics, and finite-element methods. A NEP model for representative $\beta$-$Ga_2O_3$/cBAs heterostructures is constructed, and the interfacial thermal conductance is then calculated via NEMD simulations. Exceptionally high $G$ values around 800 $MWm^{-2}K^{-1}$ are obtained at room temperature, which is primarily attributed to the similar Debye temperatures of $\beta$-$Ga_2O_3$ and cBAs and their well-matched phonon density of states, as confirmed by the DFT and MD analysis. The effects of the atomic bonding and interfacial modes are also analyzed. Furthermore, device-level thermal modeling shows that the use of cBAs substrate consistently leads to a lower temperature rise compared to other high-$\kappa$ materials such as diamond, 3C-SiC, and Si. In addition to its ultrahigh thermal conductivity, this work further reveals the potential of cBAs to form interfaces of ultrahigh thermal conductance with $\beta$-$Ga_2O_3$ and other important semiconductors. Our findings highlight cBAs as an ideal material for the thermal management of electronics, and will hopefully facilitate future engineering of interfacial thermal transport at the atomic scale.

**DATA AVAILABILITY**

The data that support the findings of this study are available from the corresponding authors upon reasonable request.

**ACKNOWLEDGMENTS**

We acknowledge Zheyong Fan for the helpful discussion on the constructions of NEP model. This work was supported by the National Key R & D Project from the Ministry of Science and Technology of China (Grant No. 2022YFA1203100) and the High-performance Computing Platform of Peking University. B.S. acknowledges support from the New Cornerstone Science Foundation through the XPLORER PRIZE.

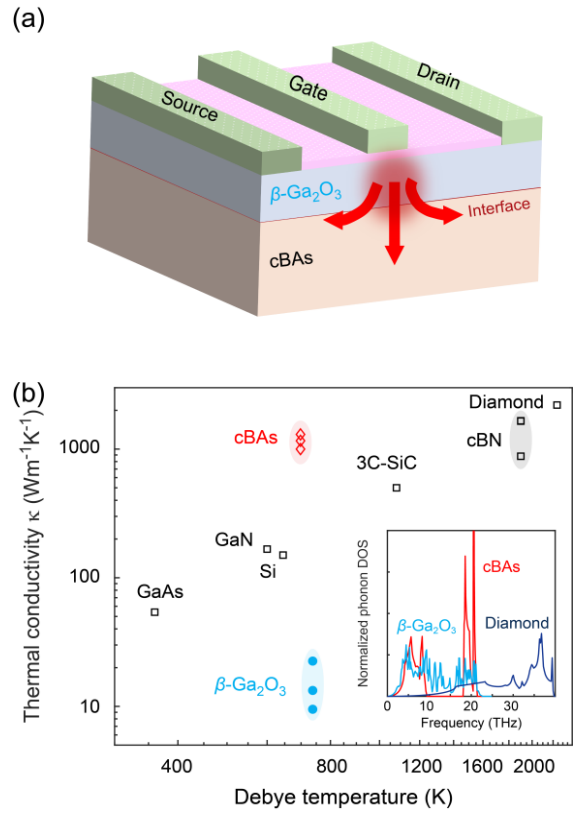

**Figure 1. Thermal management of *β*-Ga$_2$O$_3$ electronics.** (a) Schematic illustration of heat dissipation in a *β*-Ga$_2$O$_3$ device. (b) Experimentally measured thermal conductivities of representative semiconductors with respect to their Debye temperatures. The red, black, and blue shades highlight values respectively for high-quality cBAs [20-22], natural and isotope-enriched cBN [63], and *β*-Ga$_2$O$_3$ along the [100], [010], and [001] crystallographic directions [12]. Inset shows the DFT-calculated normalized phonon DOS of *β*-Ga$_2$O$_3$, cBAs, and diamond.

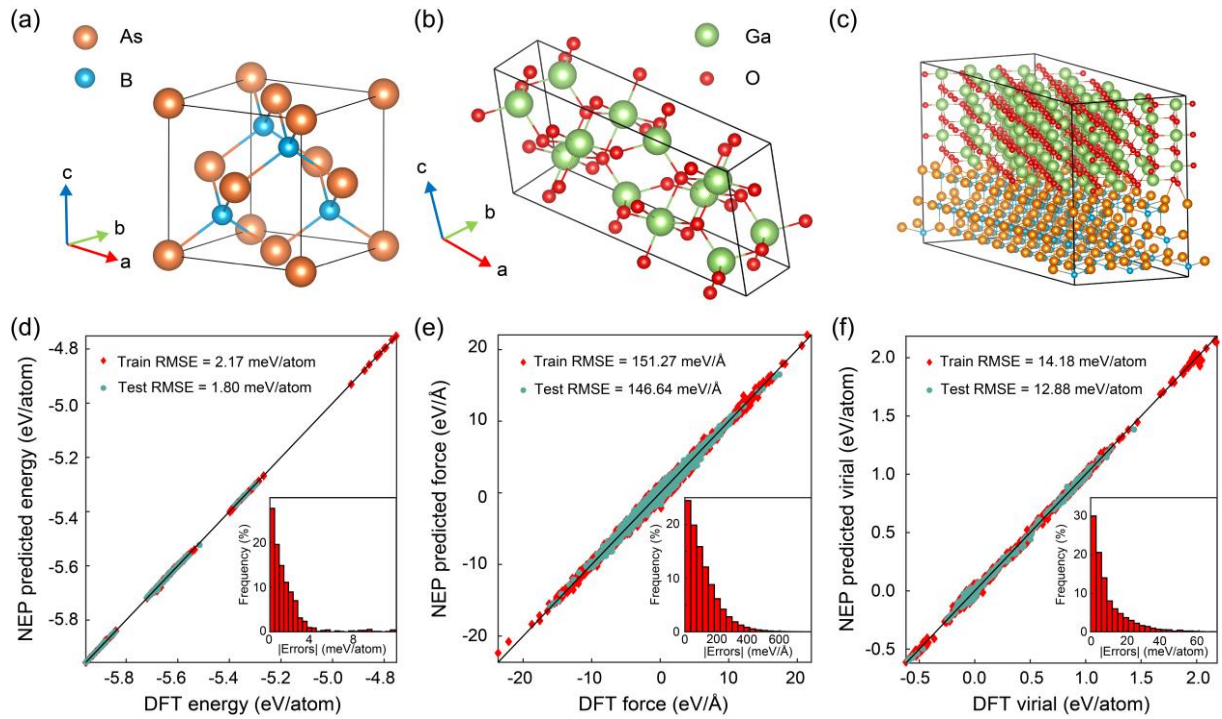

**Figure 2. Atomic structures and machine-learned potentials.** (a) The conventional unit cell of cBAs. (b) The conventional unit cell of *β*-$Ga_2O_3$. (c) The *β*-$Ga_2O_3$/cBAs heterostructure with expanded cells on both sides of the interface. (d)-(f) Parity plots for the energy, forces, and virial stresses. Insets show the probability distributions of the absolute errors, defined as the difference between the DFT and NEP predictions.

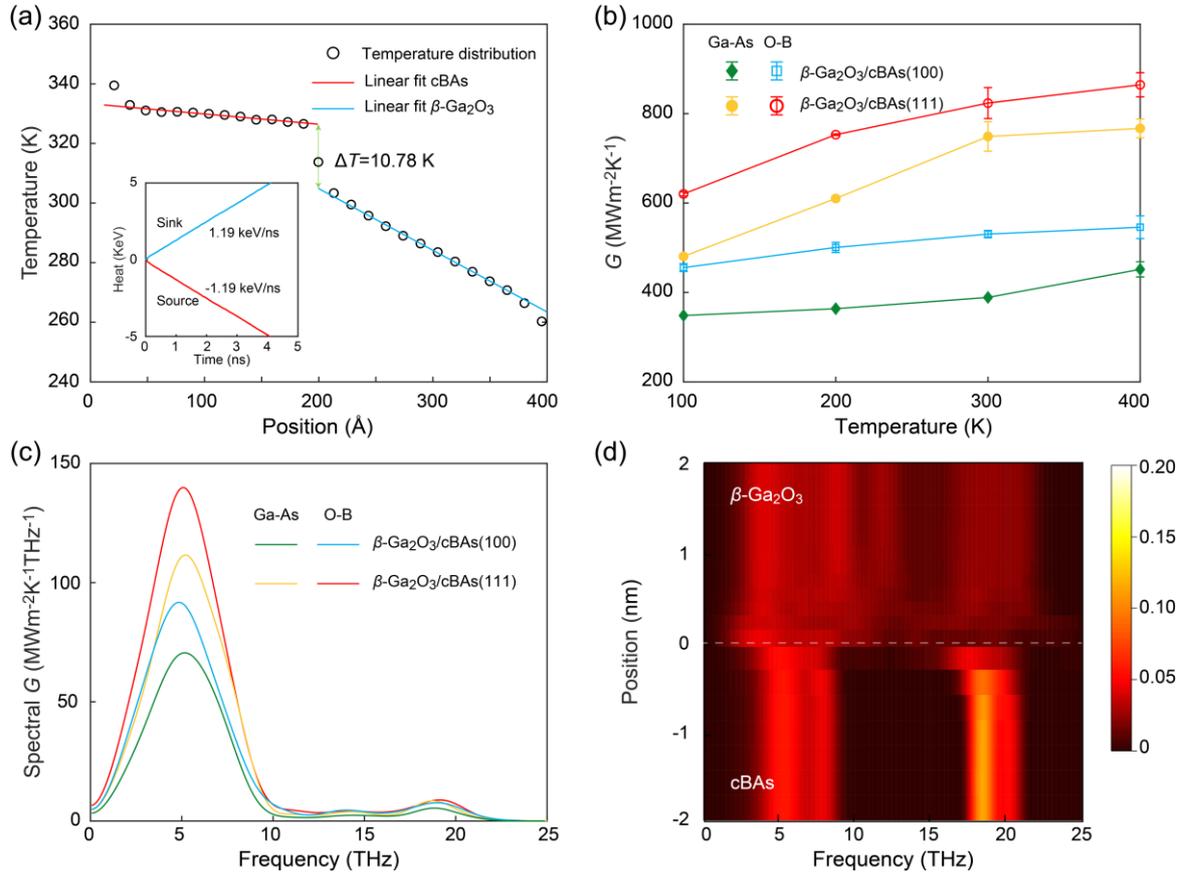

**Figure 3. Interfacial thermal conductance and phonon density of states.** (a) Steady-state temperature distribution in the *β*-Ga$_2$O$_3$(010)/cBAs(111) heterostructure. The red and blue lines are the linear fittings to the temperature profiles within cBAs and *β*-Ga$_2$O$_3$, respectively. Inset shows the accumulated heat fluxes in the thermostats, with the slopes labeled. (b) Calculated interfacial thermal conductance of the *β*-Ga$_2$O$_3$/cBAs interface as a function of temperature. (c) Spectral interfacial thermal conductance at 300 K. (d) Normalized phonon DOS near the interface. The dashed white line marks the interface.

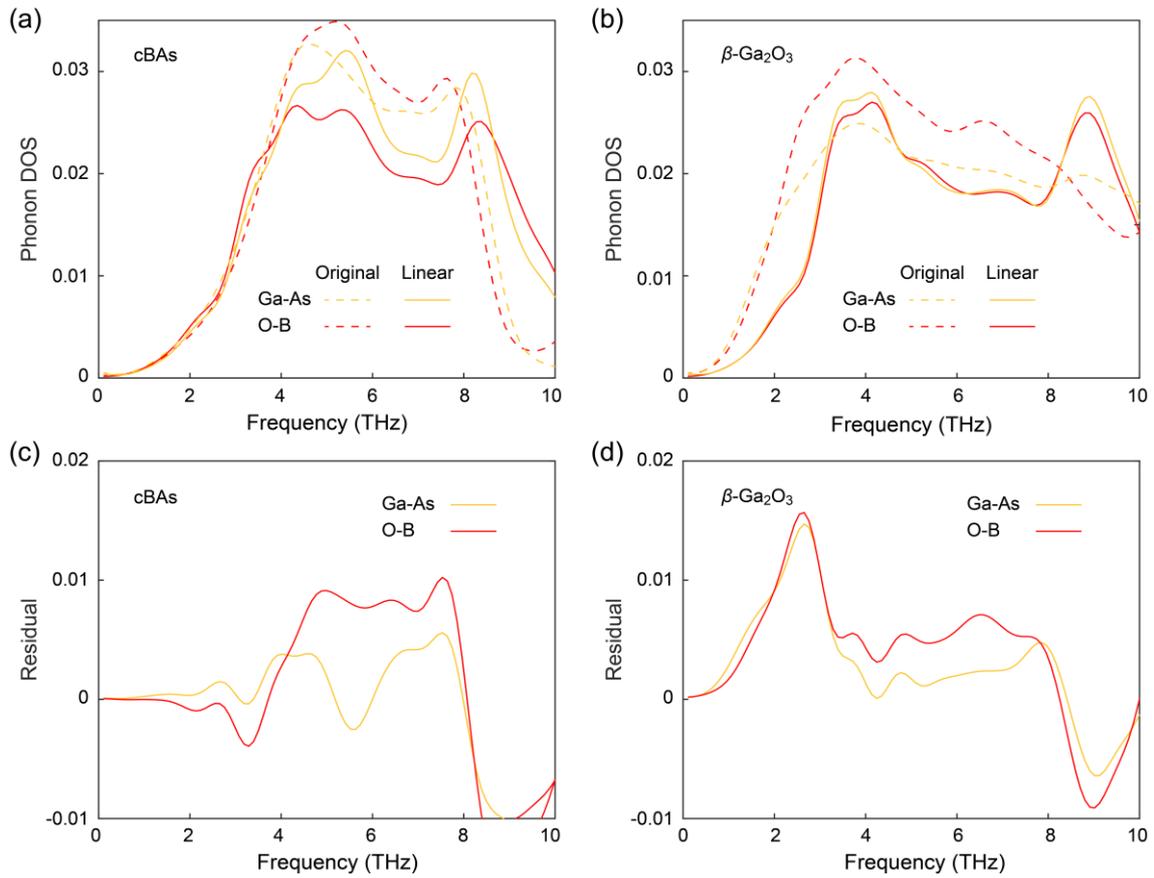

**Figure 4. Analysis of the interfacial vibrational states.** Normalized phonon DOS near the interface for the (a) cBAs and (b) *β*-Ga$_2$O$_3$ side together with the least-square fittings based on linear combinations of the bulk DOS, which is chosen from regions 10 nm away from the interface. Residuals of the fittings for the (c) cBAs and (d) *β*-Ga$_2$O$_3$ side. The calculations are performed for the *β*-Ga$_2$O$_3$(010)/cBAs(111) interface with Ga-As and O-B bonding.

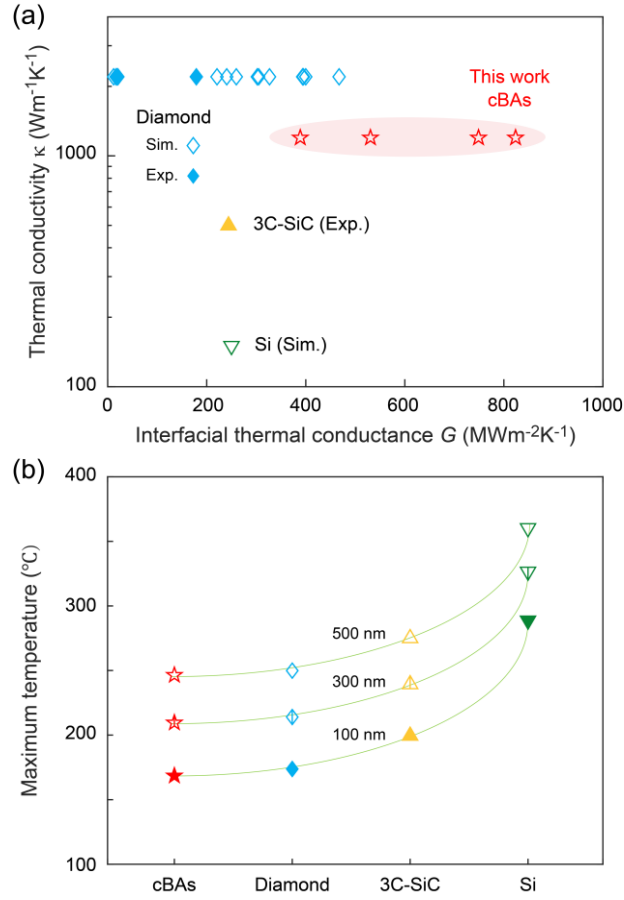

**Figure 5. Finite-element simulations of a *β*-Ga$_2$O$_3$ device on different substrates.** (a) Comparison of the room-temperature interfacial thermal conductance between *β*-Ga$_2$O$_3$ and a few representative high-*κ* materials. Both calculated (filled symbols) and measured (empty symbols) values are included. The data points for *β*-Ga$_2$O$_3$ on diamond, 3C-SiC, and Si are taken from Refs. [24, 58-60], Ref. [61], and Ref. [62], respectively. (b) Maximum *β*-Ga$_2$O$_3$ device temperature using cBAs, diamond, 3C-SiC, and Si as the substrate. Three different thicknesses (500 nm, 300 nm, and 100 nm) of the *β*-Ga$_2$O$_3$ layer are considered.

# REFERENCES

[1] J. C. Zhang, P. F. Dong, K. Dang, Y. N. Zhang, Q. L. Yan, H. Xiang, J. Su, Z. H. Liu, M. W. Si, J. C. Gao, *et al.*, Ultra-wide bandgap semiconductor $Ga_2O_3$ power diodes, Nat. Commun. **13** (1), 3900 (2022).

[2] A. J. Green, K. D. Chabak, M. Baldini, N. Moser, R. Gilbert, R. C. Fitch, G. Wagner, Z. Galazka, J. McCandless, A. Crespo, *et al.*, *β*-$Ga_2O_3$ MOSFETs for radio frequency operation, IEEE Electron Device Lett. **38** (6), 790-793 (2017).

[3] Y. H. Wang, H. R. Li, J. Cao, J. Y. Shen, Q. Y. Zhang, Y. T. Yang, Z. G. Dong, T. H. Zhou, Y. Zhang, W. H. Tang, *et al.*, Ultrahigh gain solar blind avalanche photodetector using an amorphous $Ga_2O_3$-based heterojunction, Acs Nano **15** (10), 16654-16663 (2021).

[4] J. L. Zhao, X. R. Huang, Y. H. Yin, Y. K. Liao, H. W. Mo, Q. K. Qian, Y. Z. Guo, X. L. Chen, Z. F. Zhang and M. Y. Hua, Two-dimensional gallium oxide monolayer for gas-sensing application, J Phys. Chem. Lett. **12** (24), 5813-5820 (2021).

[5] S. J. Pearton, J. Yang, I. Patrick H. Cary, F. Ren, J. Kim, M. J. Tadjer and M. A. Mastro, A review of $Ga_2O_3$ materials, processing, and devices, Appl. Phys. Rev. **5**, 011301 (2018).

[6] M. J. Tadjer, Toward gallium oxide power electronics, Science **378** (6621), 724-725 (2022).

[7] B. J. Baliga, Power semiconductor-device figure of merit for high-frequency applications, IEEE Electron Device Lett. **10** (10), 455-457 (1989).

[8] K. Sasaki, M. Higashiwaki, A. Kuramata, T. Masui and S. Yamakoshi, MBE grown $Ga_2O_3$ and its power device applications, J Cryst. Growth **378**, 591-595 (2013).

[9] Saurav Roy, Arkka Bhattacharyya, Praneeth Ranga, Heather Splawn, Jacob Leach and S. Krishnamoorthy, High-k oxide field-plated vertical (001) *β*-$Ga_2O_3$ Schottky barrier diode with Baliga's figure of merit over 1 $GW/cm^2$, IEEE Electron Device Lett. **42** (8), 1140-1143 (2021).

[10] A. Azarov, J. G. Fernandez, J. Zhao, F. Djurabekova, H. He, R. He, O. Prytz, L. Vines, U.

Bektas, P. Chekhonin, *et al.*, Universal radiation tolerant semiconductor, Nat. Commun. **14** (1), 4855 (2023).

[11] M. Slomski, N. Blumenschein, P. P. Paskov, J. F. Muth and T. Paskova, Anisotropic thermal conductivity of *β*-Ga$_2$O$_3$ at elevated temperatures: Effect of Sn and Fe dopants, J. Appl. Phys **121**, 235104 (2017).

[12] P. Jiang, X. Qian, X. Li and R. Yang, Three-dimensional anisotropic thermal conductivity tensor of single crystalline *β*-Ga$_2$O$_3$, Appl. Phys. Lett. **113**, 232105 (2018).

[13] Z. Guo, A. Verma, X. Wu, F. Sun, A. Hickman, T. Masui, A. Kuramata, M. Higashiwaki, D. Jena and T. Luo, Anisotropic thermal conductivity in single crystal *β*-gallium oxide, Appl. Phys. Lett. **106** (11), 111909 (2015).

[14] M. Handwerg, R. Mitdank, Z. Galazka and S. F. Fischer, Temperature-dependent thermal conductivity in Mg-doped and undoped *β*-Ga$_2$O$_3$ bulk-crystals, Semicond. Sci. Technol **30** (2), 024006 (2015).

[15] Z. Galazka, K. Irmscher, R. Uecker, R. Bertram, M. Pietsch, A. Kwasniewski, M. Naumann, T. Schulz, R. Schewski, D. Klimm, *et al.*, On the bulk *β*-Ga$_2$O$_3$ single crystals grown by the Czochralski method, J Cryst. Growth **404**, 184-191 (2014).

[16] X. Wang, J. Yang, P. Ying, Z. Fan, J. Zhang and H. Sun, Dissimilar thermal transport properties in *κ*-Ga$_2$O$_3$ and *β*-Ga$_2$O$_3$ revealed by homogeneous nonequilibrium molecular dynamics simulations using machine-learned potentials, J. Appl. Phys **135**, 065104 (2024).

[17] B. Chatterjee, K. Zeng, C. D. Nordquist, U. Singisetti and S. Choi, Device-level thermal management of gallium oxide field-effect transistors, IEEE T. Comp. Pack. Man. **9** (12), 2352-2365 (2019).

[18] N. Nepal, D. S. Katzer, B. P. Downey, V. D. Wheeler, L. O. Nyakiti, D. F. Storm, M. T. Hardy, J. A. Freitas, E. N. Jin, D. Vaca, *et al.*, Heteroepitaxial growth of *β*-Ga$_2$O$_3$ films on SiC via molecular beam epitaxy, J. Vac. Sci. Technol. A **38**, 063406 (2020).

[19] W. H. Xu, Y. B. Wang, T. G. You, X. Ou, G. Q. Han, H. D. Hu, S. B. Zhang, F. W. Mu, T. Suga, Y. H. Zhang, *et al.*, First demonstration of wafer scale heterogeneous integration of Ga$_2$O$_3$ MOSFETs on SiC and Si substrates by ion-cutting process, *2019 IEEE International Electron Devices Meeting (IEDM)*, IEEE (2019).

[20] F. Tian, B. Song, X. Chen, N. K. Ravichandran, Y. C. Lv, K. Chen, S. Sullivan, J. Kim, Y. Y. Zhou, T.-H. Liu, *et al.*, Unusual high thermal conductivity in boron arsenide bulk crystals, Science **361** (6402), 582–585 (2018).

[21] S. Li, Q. Zheng, Y. Lv, X. Liu, X. Wang, P. S. E. Y. Huang, D. G. Cahill and B. Lv, High thermal conductivity in cubic boron arsenide crystals, Science **361** (6402), 579-581 (2018).

[22] J. Kang, M. Li, H. Wu, H. Nguyen and Y. Hu, Experimental observation of high thermal conductivity in boron arsenide, Science **361** (6402), 575-578 (2018).

[23] K. Yaddanapudi, Ab initio calculations of the thermal properties of boron arsenide, Comp. Mater. Sci. **184**, 109887 (2020).

[24] A. Petkov, A. Mishra, J. W. Pomeroy and M. Kuball, Molecular dynamics study of thermal transport across Ga$_2$O$_3$–diamond interfaces, Appl. Phys. Lett. **122** (3), 031602 (2023).

[25] P. E. Blochl, Projector augmented-wave method, Phys. Rev. B **50** (24), 17953-17979 (1994).

[26] J. P. Perdew, K. Burke and M. Ernzerhof, Generalized gradient approximation made simple, Phys. Rev. Lett. **77** (18), 3865-3868 (1996).

[27] A. Togo, L. Chaput, T. Tadano and I. Tanaka, Implementation strategies in phonopy and phono3py, J. Phys.-Condens. Mat. **35** (35), 353001 (2023).

[28] See the Supplemental Material for the computational methods, atomic models, MLP validations, NEMD results, FEM, MLP dataset, and parameters for FD and FEM simulations, which includes Refs. [25, 26, 35-37, 44].

[29] Z. W. Ding, Q. X. Pei, J. W. Jiang, W. X. Huang and Y. W. Zhang, Interfacial thermal

conductance in graphene/MoS$_2$ heterostructures, Carbon **96**, 888-896 (2016).

[30] X. Wu and Q. Han, Semidefective graphene/h-BN in-plane heterostructures: Enhancing interface thermal conductance by topological defects, J. Phys. Chem. C **125** (4), 2748-2760 (2021).

[31] G. Chen, *Nanoscale Energy Transport and Conversion: A Parallel Treatment of Electrons, Molecules, Phonons, and Photons* (Oxford University Press, Oxford, 2005).

[32] F. Yang, W. Zhou, Z. Zhang, X. Huang, J. Zhang, N. Liang, W. Yan, Y. Wang, M. Ding, Q. Guo*, et al.*, Phonon heat conduction across slippery interfaces in twisted graphite. https://doi.org/10.48550/arXiv.2406.03758, 2024.

[33] J. Wu, E. Zhou, A. Huang, H. Zhang, M. Hu and G. Qin, Deep-potential enabled multiscale simulation of gallium nitride devices on boron arsenide cooling substrates, Nat. Commun. **15** (1), 2540 (2024).

[34] J. S. Kang, M. Li, H. Wu, H. Nguyen, T. Aoki and Y. Hu, Integration of boron arsenide cooling substrates into gallium nitride devices, Nat. Electron. **4** (6), 416-423 (2021).

[35] Z. Fan, Z. Zeng, C. Zhang, Y. Wang, K. Song, H. Dong, Y. Chen and T. Nissila, Neuroevolution machine learning potentials: Combining high accuracy and low cost in atomistic simulations and application to heat transport, Phys. Rev. B **104** (10), 104309 (2021).

[36] H. Dong, Y. Shi, P. Ying, K. Xu, T. Liang, Y. Wang, Z. Zeng, X. Wu, W. Zhou, S. Xiong*, et al.*, Molecular dynamics simulations of heat transport using machine-learned potentials: A mini-review and tutorial on GPUMD with neuroevolution potentials, J. Appl. Phys **135** (16), 161101 (2024).

[37] Z. Y. Fan, W. Chen, V. Vierimaa and A. Harju, Efficient molecular dynamics simulations with many-body potentials on graphics processing units, Comput. Phys. Commun. **218**, 10-16 (2017).

[38] W. J. Zhou and B. Song, Isotope effect on four-phonon interaction and lattice thermal transport: An atomistic study of lithium hydride, Phys. Rev. B **110** (20), 205202 (2024).

[39] P. Ying, W. Zhou, L. Svensson, E. Fransson, F. Eriksson, K. Xu, T. Liang, B. Song, S. Chen, P. Erhart, *et al.*, Highly efficient path-integral molecular dynamics simulations with GPUMD using neuroevolution potentials: Case studies on thermal properties of materials. https://doi.org/10.48550/arXiv.2409.04430, 2024.

[40] Z. Feng, Y. Cai, Z. Li, Z. Hu, Y. Zhang, X. Lu, X. Kang, J. Ning, C. Zhang, Q. Feng, *et al.*, Design and fabrication of field-plated normally off *β*-$Ga_2O_3$ MOSFET with laminated-ferroelectric charge storage gate for high power application, Appl. Phys. Lett. **116**, 243503 (2020).

[41] M. H. Wong, K. Sasaki, A. Kuramata, S. Yamakoshi and M. Higashiwaki, Field-plated GaO MOSFETs with a breakdown voltage of over 750 V, IEEE Electron Device Lett. **37** (2), 212-215 (2016).

[42] M. Higashiwaki, K. Sasaki, A. Kuramata, T. Masui and S. Yamakoshi, Gallium oxide ($Ga_2O_3$) metal-semiconductor field-effect transistors on single-crystal *β*-$Ga_2O_3$ (010) substrates, Appl. Phys. Lett. **100** (1), 013504 (2012).

[43] Z. Zhang, R. Cao, C. Wang, H. B. Li, H. Dong, W. H. Wang, F. Lu, Y. Cheng, X. Xie, H. Liu, *et al.*, GaN as an interfacial passivation layer: Tuning band offset and removing fermi level pinning for III-V MOS devices, ACS Appl. Mater. Interfaces **7** (9), 5141-5149 (2015).

[44] W. Zhou, N. Liang, X. Wu, S. Xiong, Z. Fan and B. Song, Insight into the effect of force error on the thermal conductivity from machine-learned potentials, Mater. Today Phys. **50**, 101638 (2025).

[45] J. Zhang, H. Zhang, J. Wu, X. Qian, B. Song, C. Lin, T.-H. Liu and R. Yang, Vacancy-induced phonon localization in boron arsenide using a unified neural network interatomic potential, Cell Rep. Phys. Sci. **5** (1), 101760 (2024).

[46] X. Wu, W. Zhou, H. Dong, P. Ying, Y. Wang, B. Song, Z. Fan and S. Xiong, Correcting

force error-induced underestimation of lattice thermal conductivity in machine learning molecular dynamics, J. Chem. Phys. **161**, 014103 (2024).

[47] J. Chen, X. F. Xu, J. Zhou and B. W. Li, Interfacial thermal resistance: Past, present, and future, Rev. Mod. Phys. **94** (2), 025002 (2022).

[48] P. L. Kapitza, The study of heat transfer in helium II, J Phys-Ussr **4** (1-6), 181-210 (1941).

[49] Q. Li, F. Liu, S. Hu, H. Song, S. Yang, H. Jiang, T. Wang, Y. K. Koh, C. Zhao, F. Kang, *et al.*, Inelastic phonon transport across atomically sharp metal/semiconductor interfaces, Nat. Commun. **13** (1), 4901 (2022).

[50] L. R. D., *CRC Handbook of Chemistry and Physics* (Boca Raton, CRC press, 2004).

[51] Z. Fan, H. Dong, A. Harju and T. Ala-Nissila, Homogeneous nonequilibrium molecular dynamics method for heat transport and spectral decomposition with many-body potentials, Phys. Rev. B **99** (6), 064308 (2019).

[52] A. J. Gabourie, Z. Y. Fan, T. Ala-Nissila and E. Pop, Spectral decomposition of thermal conductivity: Comparing velocity decomposition methods in homogeneous molecular dynamics simulations, Phys. Rev. B **103** (20), 205421 (2021).

[53] W. Zhou, Y. Dai, J. Zhang, B. Song, T.-H. Liu and R. Yang, Effect of four-phonon interaction on phonon thermal conductivity and mean-free-path spectrum of high-temperature phase SnSe, Appl. Phys. Lett. **121** (11), 112202 (2022).

[54] Z. Cheng, R. Li, X. Yan, G. Jernigan, J. Shi, M. E. Liao, N. J. Hines, C. A. Gadre, J. C. Idrobo, E. Lee, *et al.*, Experimental observation of localized interfacial phonon modes, Nat. Commun. **12** (1), 6901 (2021).

[55] Y. H. Li, R. S. Qi, R. C. Shi, J. N. Hu, Z. T. Liu, Y. W. Sun, M. Q. Li, N. Li, C. L. Song, L. Wang, *et al.*, Atomic-scale probing of heterointerface phonon bridges in nitride semiconductor, P. Natl. Acad. Sci. USA **119** (8), e2117027119 (2022).

[56] R. S. Qi, R. C. Shi, Y. H. Li, Y. W. Sun, M. Wu, N. Li, J. L. Du, K. H. Liu, C. L. Chen, J.

Chen, *et al.*, Measuring phonon dispersion at an interface, Nature **599** (7885), 399-403 (2021).

[57] K. Gordiz and A. Henry, Phonon transport at interfaces: Determining the correct modes of vibration, J. Appl. Phys **119** (1), 015101 (2016).

[58] Z. Cheng, V. D. Wheeler, T. Bai, J. Shi, M. J. Tadjer, T. Feygelson, K. D. Hobart, M. S. Goorsky and S. Graham, Integration of polycrystalline $Ga_2O_3$ on diamond for thermal management, Appl. Phys. Lett. **116** (6), 062105 (2020).

[59] Z. Cheng, L. Yates, J. Shi, M. J. Tadjer, K. D. Hobart and S. Graham, Thermal conductance across $β$-$Ga_2O_3$-diamond van der Waals heterogeneous interfaces, APL Mater. **7** (3), 031118 (2019).

[60] Z. Sun, D. Zhang, Z. Qi, Q. Wang, X. Sun, K. Liang, F. Dong, Y. Zhao, D. Zou, L. Li, *et al.*, Insight into interfacial heat transfer of $β$-$Ga_2O_3$/diamond heterostructures via the machine learning potential, ACS Appl. Mater. Interfaces **16** (24), 31666-31676 (2024).

[61] J. Liang, H. Nagai, Z. Cheng, K. Kawamura, Y. Shimizu, Y. Ohno, Y. Sakaida, H. Uratani, H. Yoshida, Y. Nagai, *et al.*, Selective direct bonding of high thermal conductivity 3C-SiC film to $β$-$Ga_2O_3$ for top-side heat extraction. https://doi.org/10.48550/arXiv.2209.05669, 2022.

[62] Y. Li, F. Sun and Y. Feng, Thermal boundary conductance in heterogeneous integration between $β$-$Ga_2O_3$ and semiconductors, Ceram. Int. **50** (11), 18787-18796 (2024).

[63] K. Chen, B. Song, N. K. Ravichandran, Q. Zheng, X. Chen, H. Lee, H. Sun, S. Li, G. A. G. U. Gamage, F. Tian, *et al.*, Ultrahigh thermal conductivity in isotope-enriched cubic boron nitride, Science **367**, 555-529 (2020).